\documentclass[10pt]{article}
\usepackage{amsmath}
\usepackage{amssymb}
\usepackage{graphicx}
\usepackage{cite}
\usepackage{color}
\usepackage[labelfont=bf,labelsep=period,justification=raggedright]{caption}

\makeatletter
\renewcommand{\@biblabel}[1]{\quad#1.}
\makeatother

\date{}

\begin{document}

\begin{flushleft}
{\Large
\textbf{Evolutionary establishment of moral and double moral standards through spatial interactions}
}\sffamily
\\[3mm]
\textbf{Dirk Helbing$^{1,2,3,\ast}$,
Attila Szolnoki$^{4}$,
Matja{\v z} Perc$^{5}$,
Gy{\"o}rgy Szab{\'o}$^{4}$}
\\[2mm]
{\bf 1} ETH Zurich, CLU E1, Clausiusstr. 50, 8092 Zurich, Switzerland,
{\bf 2} Santa Fe Institute, 1399 Hyde Park Road, Santa Fe, NM 87501, USA,
{\bf 3} Collegium Budapest - Institute for Advanced Study,
{\bf 4} Research Institute for Technical Physics and Materials Science, P.O. Box 49, H-1525 Budapest, Hungary,
{\bf 5} Faculty of Natural Sciences and Mathematics, University of Maribor, Koro{\v s}ka cesta 160, SI-2000 Maribor, Slovenia
\\
$\ast$ E-mail: dhelbing@ethz.ch
\end{flushleft}
\sffamily
\section*{Abstract}
Situations where individuals have to contribute to joint efforts or share scarce resources are ubiquitous. Yet, without proper mechanisms to ensure cooperation, the evolutionary pressure to maximize individual success tends to create a tragedy of the commons (such as over-fishing or the destruction of our environment). This contribution addresses a number of related puzzles of human behavior with an evolutionary game theoretical approach as it has been successfully used to explain the behavior of other biological species many times, from bacteria to vertebrates. Our agent-based model distinguishes individuals applying four different behavioral strategies: non-cooperative individuals (``defectors''), cooperative individuals abstaining from punishment efforts (called ``cooperators'' or ``second-order free-riders''), cooperators who punish non-cooperative behavior (``moralists''), and defectors, who punish other defectors despite being non-cooperative themselves (``immoralists''). By considering spatial interactions with neighboring individuals, our model reveals several interesting effects: First, moralists can fully eliminate cooperators. This spreading of punishing behavior requires a segregation of behavioral strategies and solves the ``second-order free-rider problem''.  Second, the system behavior changes its character significantly even after very long times (``who laughs last laughs best effect''). Third, the presence of a number of defectors can largely accelerate the victory of moralists over non-punishing cooperators. Forth, in order to succeed, moralists may profit from immoralists in a way that appears like an ``unholy collaboration''. Our findings suggest that the consideration of punishment strategies allows to understand the establishment and spreading of ``moral behavior'' by means of game-theoretical concepts. This demonstrates that quantitative biological modeling approaches are powerful even in domains that have been addressed with non-mathematical concepts so far. The complex dynamics of certain social behaviors becomes understandable as result of an evolutionary competition between different behavioral strategies.

\section*{Author Summary}
Why do friends spontaneously come up with mutually accepted rules, cooperation, and solidarity, while the creation of shared moral standards often fails in large communities? In a ``global village'', where everybody may interact with anybody else, it is not worthwhile to punish people who cheat. Moralists (cooperative individuals who undertake punishment efforts) disappear because of their disadvantage compared to cooperators who do not punish (so-called ``second-order free-riders''). However, cooperators are exploited by free-riders. This creates a ``tragedy of the commons'', where everybody is uncooperative in the end. Yet, when people interact with friends or local neighbors, as most people do, moralists can escape the direct competition with non-punishing cooperators by separating from them. Moreover, in the competition with free-riders, moralists can defend their interests better than non-punishing cooperators. Therefore, while seriously depleted in the beginning, moralists can finally spread all over the world (``who laughs last laughs best effect''). Strikingly, the presence of a few non-cooperative individuals (``deviant behavior'') can accelerate the victory of moralists. In order to spread, moralists may also form an ``unholy cooperation'' with people having double moral standards, i.e. free-riders who punish non-cooperative behavior, while being uncooperative themselves.

\section*{Introduction}
Public goods such as environmental resources or social benefits are particularly prone to exploitation by  non-cooperative individuals (``defectors''), who try to increase their benefit at the expense of fair contributors or users, the ``cooperators''. This implies a tragedy of commons \cite{1}. It was proposed that costly punishment of non-cooperative individuals can establish cooperation in public goods dilemmas \cite{2,3,4,5,6,7,SigHau}, and it is effective indeed \cite{8,9,10}. Nonetheless, why would cooperators choose to punish defectors at a personal cost \cite{Yama,11,12}? One would expect that evolutionary pressure should eventually eliminate such ``moralists'' due to their extra costs compared to ``second-order free-riders'' (i.e. cooperators, who do not punish). These, however should finally be defeated by ``free-riders'' (defectors). To overcome this problem \cite{13,14}, it was proposed that cooperators who punish defectors (called ``moralists'' by us) would survive through indirect reciprocity \cite{15}, reputation effects \cite{16} or the possibility to abstain from the joint enterprize \cite{17,18,19} by ``volunteering'' \cite{20,21}. Without such mechanisms, cooperators who punish will usually vanish. Surprisingly, however, the second-order free-rider problem is naturally resolved, without assuming additional mechanisms, if spatial or network interactions are considered. This will be shown in the following.

In order to study the conditions for the disappearance of non-punishing cooperators and defectors, we simulate the public goods game with costly punishment, considering two cooperative strategies (C, M) and two defective ones (D, I). For illustration, one may imagine that cooperators (C) correspond to countries trying to meet the CO$_2$ emission standards of the Kyoto protocol \cite{Mili}, and ``moralists'' (M) to cooperative countries that additionally enforce the standards by international pressure (e.g. embargoes). Defectors (D) would correspond to those countries ignoring the Kyoto protocol, and immoralists (I) to countries failing to meet the Kyoto standards, but nevertheless imposing pressure on other countries to fulfil them. According to the classical game-theoretical prediction, all countries would finally fail to meet the emission standards, but we will show that, in a spatial setting, interactions between the four strategies C, D, M, and I can promote the spreading of moralists. Other well-known public goods problems are over-fishing, the pollution of our environment, the creation of social benefit systems, or the establishment and maintenance of cultural institutions (such as a shared language, norms, values, etc.).

Our simplified game-theoretical description of such problems assumes that cooperators (C) and moralists (M) make a contribution of $1$ to the respective public good under consideration, while nothing is contributed by defectors (D) and ``immoralists'' (I), i.e. defectors who punish other defectors. The sum of all contributions is multiplied by a factor $r$ reflecting {\em synergy effects} of cooperation, and the resulting amount is equally shared among the $k+1$ interacting individuals. Moreover, moralists and immoralists impose a fine $\beta/k$ on each defecting individual (playing D or I), which produces an additional cost $\gamma/k$ per punished defector to them (see Methods for details). The division by $k$ scales for the group size, but for simplicity, the parameter $\beta$ is called the {\em punishment fine} and $\gamma$ the {\em punishment cost}.

Given the same interaction partners, an immoralist never gets a higher payoff than a defector, but does equally well in a cooperative environment. Moreover, a cooperator tends to outperform a moralist, given the interaction partners are the same. However, a cooperator can do better than a defector when the punishment fine $\beta$ is large enough.

It is known that punishment in the public goods game and similar games can promote cooperation above a certain critical threshold of the synergy factor $r$ \cite{10,sigmund_book}. Besides cooperators who punish defectors, Heckathorn considered ``full cooperators'' (moralists) and ``hypocritical cooperators'' (immoralists) \cite{Hecka}. For well-mixed interactions (where individuals interact with a representative rather than local strategy distribution), Eldakar and Wilson find that altruistic punishment (moralists) can spread, if second-order free-riders (non-punishing altruists) are excluded, and that selfish punishers (immoralists) can survive together with altruistic non-punishers (cooperators), provided that selfish nonpunishers (defectors) are sufficiently scarce \cite{Eldakar2}.
\par
Besides well-mixed interactions, some researchers have also investigated the effect of spatial interactions \cite{5,10,NakaIw,SekiNaka}, since it is known that they can support the survival or spreading of cooperators \cite{Now1} (but this is not always the case \cite{Traul,Now2}). In this way, Brandt {\it et al.} discovered a coexistence of cooperators and defectors for certain parameter combinations \cite{10}. Compared to these studies, our model assumes somewhat different replication and strategy updating rules. The main point, however, is that we have chosen long simulation times and scanned the parameter space more extensively, which revealed several new insights, for example, the possible coexistence of immoralists and moralists, even when a substantial number of defectors is present initially. When interpreting our results within the context of moral dynamics \cite{Hauser}, our main discoveries for a society facing public goods games may be summarized as follows:
\begin{enumerate}
\item {\em Victory over second-order free-riders:} Over a long enough time period, moralists fully eliminate cooperators, thereby solving the ``second-order free-rider problem''. This becomes possible by spatial segregation of the two cooperative strategies C and M, where the presence of defectors puts moralists in a advantageous position,  which eventually allows moralists to get rid of non-punishing cooperators.
\item {\em ``Who laughs last laughs best effect'':} Moralists defeat cooperators even when the defective strategies I and D are eventually eliminated, but this process is very slow. That is, the system behavior changes its character significantly even after very long times. This is the essence of the ``who laughs last laughs best effect''. The finally winning strategy can be in a miserable situation in the beginning, and its victory may take very long.
\item {\em ``Lucifer's positive side effect'':} By permanently generating a number of defectors, small mutation rates can considerably accelerate the spreading of moralists.
\item {\em ``Unholy collaboration'' of moralists with immoralists:} Under certain conditions, moralists can survive by profiting from immoralists. This actually provides the first explanation for the existence of defectors, who hypocritically punish other defectors, although they defect themselves. The occurrence of this strange behavior is well-known in reality and even experimentally confirmed \cite{Falk,Shinada}.
\end{enumerate}
These discoveries required a combination of theoretical considerations and extensive computer simulations on multiple processors over long time horizons.

\section*{Results}
For well-mixed interactions, defectors are the winners of the evolutionary competition among the four behavioral strategies C, D, M, and I \cite{23}, which implies a tragedy of the commons despite punishment efforts. The reason is that cooperators (second-order free-riders) spread at the cost of moralists, while requiring them for their own survival.

Conclusions from computer simulations are strikingly different, if the assumption of well-mixed interactions is replaced by the more realistic assumption of spatial interactions. When cooperators and defectors interact in space \cite{5,10,24,25,26,sg1,sg2,sg3,sg4,sg5}, it is known that some cooperators can survive through spatial clustering \cite{27}. However, it is not clear how the spatiotemporal dynamics and the frequency of cooperation would change in the presence of moralists and immoralists. Would spatial interactions be able to promote the spreading of punishment and thereby eliminate second-order free-riders?

In order to explore this, we have scanned a large parameter space. Figure~1 shows the resulting state of the system as a function of the punishment cost $\gamma$ and punishment fine $\beta$ after a sufficiently long transient time. If the fine-to-cost ratio $\beta/\gamma$ and the synergy factor $r$ are low, defectors eliminate all other strategies.  However, for large enough fines $\beta$, cooperators and defectors are always eliminated, and moralists prevail (Fig.~1).

At larger $r$ values, when the punishment costs are moderate, we find a coexistence of moralists with defectors without any cooperators. To understand why moralists can outperform cooperators despite additional punishment costs, it is important to analyze the dynamics of spatial interactions. Starting with a homogeneous strategy distribution (Fig.~2a), the imitation of better-performing neighbors generates small clusters of individuals with identical strategies (Fig.~2b). ``Immoralists'' die out quickly, while cooperators and moralists form separate clusters in a sea of defectors (Fig.~2c). The further development is determined by the interactions at the interfaces between clusters of different strategies (Figs.~2d--f). In the presence of defectors, the fate of moralists is not decided by a {\em direct} competition with cooperators, but rather by the success of both cooperative strategies against invasion attempts by defectors. If the $\beta/\gamma$-ratio is appropriate, moralists respond better to defectors than cooperators. Indeed, moralists can spread so successfully in the presence of defectors that areas lost by cooperators are quickly occupied by moralists (supplementary Video S1). This indirect territorial battle ultimately leads to the extinction of cooperators (Fig.~2f), thus resolving the second-order free-rider problem.
\par
In conclusion, the presence of some {\em conventional} free-riders (defectors) supports the elimination of {\em second-order} free-riders. However, if the fine-to-cost ratio is high, defectors are eliminated after some time. Then, the final struggle between moralists and cooperators takes such a long time that cooperators and moralists seem to coexist in a stable way. Nevertheless, a very slow coarsening of clusters is revealed, when simulating over extremely many iterations. This process is finally won by moralists, as they are in the majority by the time the defectors disappear, while they happen to be in the minority during the first stage of the simulation (see Fig. 2). We call this the ``who laughs last laughs best effect''. Since the payoffs of cooperators and moralists are identical in the absence of other strategies, the underlying coarsening dynamics is expected to agree with the voter model \cite{Dornic}.
\par
Note that there is always a punishment fine $\beta$, for which moralists can outcompete all other strategies. The higher the synergy factor $r$, the lower the $\beta/\gamma$-ratio required to reach the prevalence of moralists. Yet, for larger values of $r$, the system behavior also becomes richer, and there are areas for small fines or high punishment costs, where clusters with different strategies can coexist (see Figs.~1b--d). For example, we observe the coexistence of clusters of moralists and defectors (see Fig.~2 and supplementary Video S1) or of cooperators and defectors (see supplementary Video S2).
\par
Finally, for low punishment costs $\gamma$ but moderate punishment fines and synergy factors $r$ (see Fig.~1d), the survival of moralists may require the coexistence
with ``immoralists'' (see Fig.~3 and supplementary Video S3).
Such immoralists are often called ``sanctimonious'' or blamed for ``double moral standards'', as they defect themselves, while enforcing the cooperation of others (for the purpose of exploitation). This is actually the main obstacle for the spreading of immoralists, as they have to pay punishment costs, while suffering from punishment fines as well. Therefore, immoralists need small punishment costs $\gamma$ to survive. As cooperators die out quickly for moderate values of $r$,
the survival of immoralists depends on the existence of moralists they can exploit, otherwise they cannot outperform defectors. Conversely, moralists benefit from immoralists by supporting the punishment of defectors. Note, however, that this mutually profitable interaction between moralists and immoralists, which appears like an ``unholy collaboration'', is fragile: If $\beta$ is increased, immoralists suffer from fines, and if $\gamma$ is increased, punishing becomes too costly. In both cases, immoralists die out, and the coexistence of moralists and immoralists breaks down. Despite this fragility, ``hypocritical'' defectors, who punish other defectors, are known to occur in reality. Their existence has even been found in experiments \cite{Falk,Shinada}. Here, we have revealed conditions for their occurrence.

\section*{Discussion}
In summary, the second-order free-rider problem finds a natural and simple explanation, without requiring additional assumptions, if the local nature of most social interactions is taken into account and punishment efforts are large enough. In fact, the presence of spatial interactions can change the system behavior so dramatically that we do not find the dominance of free-riders (defectors) as in the case of well-mixed interactions, but a prevalence of moralists via a ``who laughs last laughs best'' effect (Fig.~2). Moralists can escape disadvantageous kinds of competition with cooperators by spatial segregation. However,
their triumph over all the other strategies requires the temporary presence of defectors, who diminish the cooperators (second-order free-riders). Finally, moralists can take over, as they have reached a superiority over cooperators (which is further growing) and as they can outcompete defectors (conventional free-riders).
\par
Our findings stress how crucial spatial or network interactions in social systems are. Their consideration  gives rise to a rich variety of possible dynamics and a number of continuous or discontinuous transitions between qualitatively different system behaviors. Spatial interactions can even {\em invert} the finally expected system behavior and, thereby, explain a number of challenging puzzles of social, economic, and biological systems. This includes the higher-than-expected level of cooperation in social dilemma situations, the elimination of second-order free-riders, and the formation of what looks like a collaboration between otherwise inferior strategies.
\par
By carefully scanning the parameter space, we found several possible kinds of coexistence between two strategies each:
\begin{itemize}
\item Moralists (M) and defectors (D) can coexist, when the disadvantage of cooperative behavior is not too large (i.e. the synergy factor is high enough), and if the punishment fine is sufficiently large that moralists can survive among defectors, but not large enough to get rid of them.
\item Instead of M and D, moralists (M) and immoralists (I)  coexist, when the punishment cost is small enough. The small punishment cost is needed to ensure that the disadvantage of punishing defectors (I) compared to non-punishing defectors (D) is small enough that it can be compensated by the additional punishment efforts contributed by moralists.
\item To explain the well-known coexistence of D and C \cite{10}, it is useful to remember that defectors can be crowded out by cooperators, when the synergy factor exceeds a critical value (even when punishment is not considered). Slightly below this threshold, neither cooperators nor defectors have a sufficient advantage to get rid of the other strategy, which results in a coexistence of both strategies.
\end{itemize}
Generally, a coexistence of strategies occurs, when the payoffs at the interface between clusters of different strategies are balanced. In order to understand why the coexistence is possible in a certain parameter area rather than just for an infinitely small parameter set, it is important to consider that typical cluster sizes vary with the parameter values. This also changes the typical radius of the interface between the coexisting strategies and, thereby, the typical number of neighbors applying the same strategy or a different one. In other words, a change in the shape of a cluster can partly counter-balance payoff differences between two strategies by varying the number of ``friends'' and ``enemies'' involved in the battle at the interface between spatial areas with different strategies (see Fig. \ref{add}).
\par
Finally, we would like to discuss the robustness of our observations. It is well-known that the level of cooperation  in the public goods game is highest in {\em small} groups \cite{9}.
However, we have found that moralists can crowd out non-punishing cooperators also for group sizes of $k+1=9$, 13, 21, or 25 interacting individuals, for example.
In the limiting case of {\em large} groups, where everybody interacts with everybody else, we expect the outcome of the well-mixed case, which corresponds to defection by everybody (if other mechanisms like reputation effects \cite{10} or abstaining are not considered \cite{18}). That is, the same mechanisms that can create cooperation among friends may {\em fail} to establish shared moral standards, when spatial interactions are negligible. It would therefore be interesting to study, whether the fact that interactions in the financial system are global, has contributed to the financial crisis. Typically, when social communities exceed a certain size, they need sanctioning institutions to stabilize cooperation (such as laws, an executive system, and police).
\par
Note that our principal discoveries are not expected to change substantially for spatial interactions within {\em irregular} grids (i.e. neighborhoods different from Moore neighborhoods) \cite{Flache}. In case of {\em network} interactions, we have checked that small-world or random networks lead to similar results, when the degree distribution is the same (see Fig. \ref{add1}). A {\em heterogeneous} degree distribution is even expected to {\em reduce} free-riding \cite{24} (given the average degree is the same). Finally, adding other cooperation-promoting mechanisms to our model such as direct reciprocity (a shadow of the future through repeated interactions \cite{Axelrod}), indirect reciprocity \cite{15} (trust and reputation effects \cite{10,16}), abstaining from a joint enterprize \cite{17,18,19,20,21}, or success-driven migration \cite{withYu}, will strengthen the victory of moralists over conventional and second-order free-riders.
\par
In order to test the robustness of our observations, we have also checked the effect of randomness (``noise'') originating from the possibility of strategy mutations. It is known that mutations may promote cooperation \cite{22}. According to the numerical analysis of the spatial public goods game with punishment, the introduction of rare mutations does not significantly change the final {\it outcome} of the competition between moralists and non-punishing cooperators. Second-order free-riders will always be a negligible minority in the end, if the fine-to-cost ratio and mutation rate allows moralists to spread. While a large mutation rate naturally causes a uniform distribution of strategies, a low level of strategy mutations can be even beneficial for moralists. Namely, by permanently generating a number of defectors, small mutation rates can considerably accelerate the spreading of moralists, i.e. the slow logarithmic coarsening is replaced by another kind of dynamics \cite{accelerate}. Defectors created by mutations play the same role as in the $D+M$ phase (see Figs. 1+2). They put moralists into an advantage over non-punishing cooperators, resulting in a faster spreading of the moralists (which facilitates the elimination of second-order free-riders over realistic time periods). In this way, the presence of a few ``bad guys'' (defectors) can accelerate the spreading of moral standards. Metaphorically speaking, we call this ``lucifer's positive side effect''.
\par
The current study paves the road for several interesting extensions. It is possible, for example, to study {\em antisocial} punishment \cite{anti}, considering also strategies which punish cooperators \cite{Now}. The conditions for the survival or spreading of antisocial punishers can be identified by the {\em same} methodology, but the larger number of strategies creates new phases in the parameter space. While the added complexity transcends what can be discussed here, the current study demonstrates clearly how differentiated the moral dynamics in a society facing public goods problems can be and how it depends on a variety of factors (such as the  punishment cost, punishment fine, and synergy factor). Going one step further, evolutionary game theory may even prove useful to understand how moral feelings have evolved.
\par
Furthermore, it would be interesting to investigate the {\em emergence} of punishment within the framework of a coevolutionary model \cite{darwin,ties,rev}, where both, individual strategies and punishment levels are simultaneously spread. Such a model could, for example, assume that individuals show some exploration behavior \cite{22} and stick to successful punishment levels for a long time, while they quickly abandon unsuccessful ones. In the beginning of this coevolutionary process, costly punishment would not pay off. However, after a sufficiently long time, mutually fitting punishment strategies are expected to appear in the same neighborhood by coincidence \cite{withYu}. Once an over-critical number of successful punishment strategies have appeared in some area of the simulated space, they are eventually expected to spread. The consideration of success-driven migration should strongly support this process  \cite{withYu}. Over many generations, genetic-cultural coevolution could finally inherit costly punishment as a behavioral trait, as is suggested by the mechanisms of strong reciprocity \cite{Strong}.

\section*{Methods}
We study the public goods game with punishment. Cooperative individuals (C and M) make a contribution of 1 to the public good, while defecting individuals (D and I) contribute nothing. The sum of all contributions is multiplied by $r$ and the resulting amount equally split among the $k+1$ interacting individuals. A defecting individual (D or I) suffers a fine $\beta/k$ by each punisher among the interaction partners, and each punishment requires a punisher (M or I) to spend a cost $\gamma/k$ on each defecting individual among the interaction partners. In other words, only defectors and punishing defectors (immoralists) are punished, and the overall punishment is proportional to the sum of moralists and immoralists among the $k$ neighbors. The scaling by $k$ serves to make our results comparable with models studying different groups sizes.
\par
Denoting the number of so defined cooperators, defectors, moralists, and immoralists among the $k$ interaction partners by $N_C$, $N_D$, $N_{M}$ and $N_{I}$, respectively, an individual obtains the following payoff: If it is a cooperator, it gets $P_C=r(N_C+N_{M}+1)/(k+1) - 1$, if a defector, the payoff is $P_D= r(N_C+N_{M})/(k+1) - \beta (N_{M}+N_{I}) /k$, a moralist receives $P_{M}=P_C - \gamma(N_D+N_{I})/k$, and an immoralist obtains $P_{I}= P_D - \gamma (N_D+N_{I})/k$. Our model of the spatial variant of this game studies interactions in a simple social network allowing for clustering. It assumes that individuals are distributed on a square lattice with periodic boundary conditions and play a public goods game with $k=4$ neighbors. We work with a fully occupied lattice of size $L\times L$ with $L= 200...1200$ in Fig.~1 and $L=100$ in Figs.~2--4 (the lattice size must be large enough to avoid an accidental extinction of a strategy). The initial strategies of the $L^2$ individuals are equally and uniformly distributed. Then, we perform a random sequential update. The individual at the randomly chosen location $x$ belongs to five groups. (It is the focal individual of a Moore neighborhood and a member of the Moore neighborhoods of four nearest neighbors). It plays the public goods game with the $k$ interaction partners of a group $g$, and obtains a payoff $P_x^g$ in all 5 groups it belongs to. The overall payoff is $P_x = \sum_g P_x^g$. Next, one of the four nearest neighbors is randomly chosen. Its location shall be denoted by $y$ and its overall payoff by $P_y$. This neighbor imitates the strategy of the individual at location $x$ with probability $q=1/\{1+\exp[(P_y-P_x)/K]\}$ \cite{27}. That is, individuals tend to imitate better performing strategies in their neighborhood, but sometimes deviate (due to trial-and-error behavior or mistakes) \cite{Traul}. Realistic noise levels lie between the two extremes $K\rightarrow 0$ (corresponding to unconditional imitation by the neighbor, whenever the overall payoff $P_x$ is higher than $P_y$) and $K\rightarrow \infty$ (where the strategy is copied with probability 1/2, independently of the payoffs). For the noise level $K=0.5$ chosen in our study, the evolutionary selection pressure is high enough to eventually eliminate poorly performing strategies in favor of strategies with a higher overall payoff. This implies that the resulting frequency distribution of strategies in a large enough lattice is independent of the specific initial condition after a sufficiently long transient time. Close to the separating line between M and D+M in Fig.~1,  the equilibration may require up to $10^7$ iterations (involving $L^2$ updates each).

\section*{Acknowledgments}
We acknowledge partial financial support from the EU Project QLectives and the ETH Competence Center ``Coping with Crises in Complex Socio-Economic Systems'' (CCSS) through ETH Research Grant CH1-01 08-2 (D.H.), from the Hungarian National Research Fund (grant K-73449 to A.S. and G.S.), the Bolyai Research Grant (to A.S.), the Slovenian Research Agency (grant Z1-2032-2547 to M.P.), and the Slovene-Hungarian bilateral incentive (grant BI-HU/09-10-001 to A.S., M.P. and G.S.). D.H. would like to thank for useful comments by Carlos P. Roca, Moez Draief, Stefano Balietti, Thomas Chadefaux, and Sergi Lozano.

\clearpage


\begin{figure}[!ht]
\begin{center}
\includegraphics[width=4.5in]{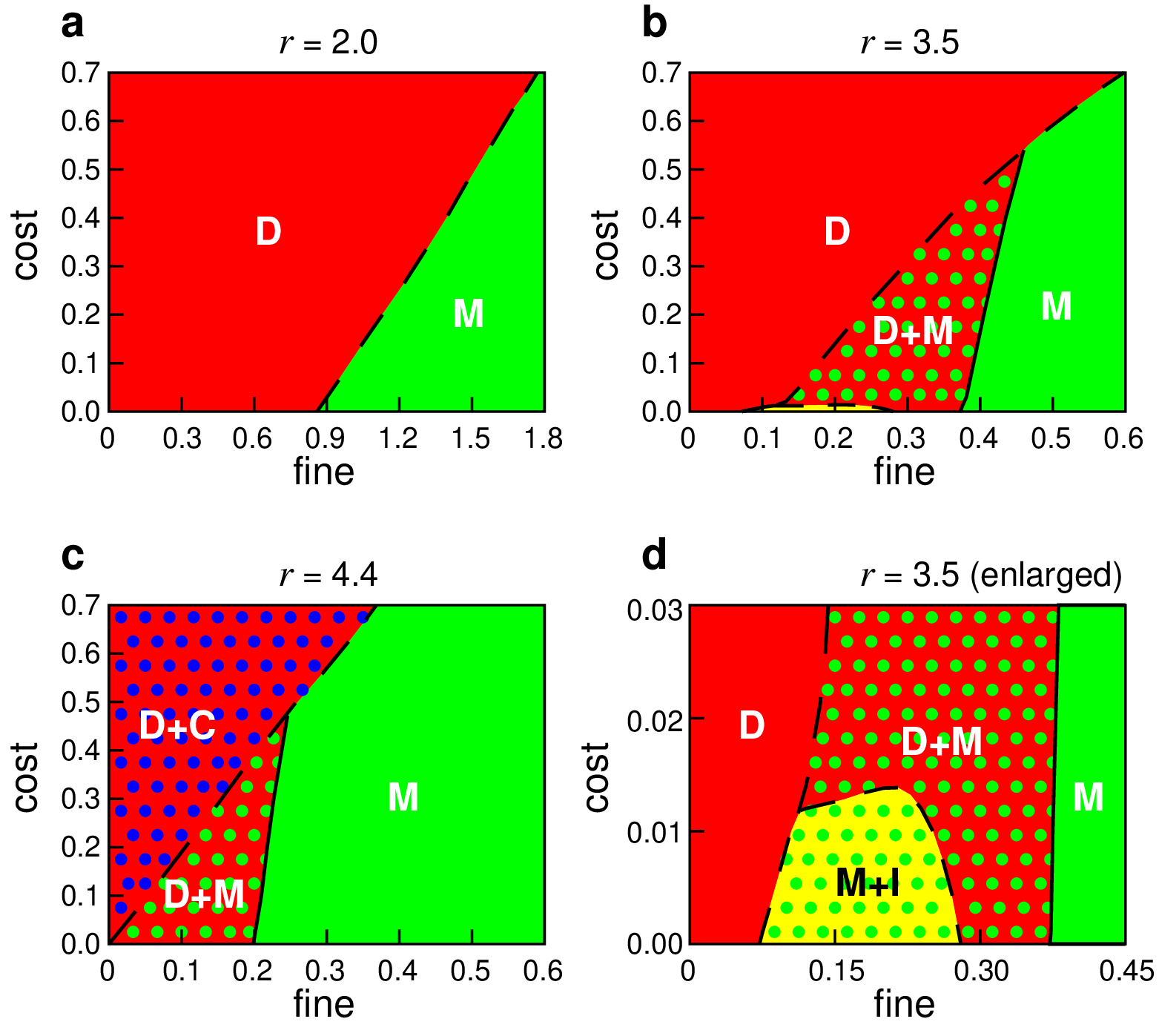}
\end{center}
\caption{{\bf Phase diagrams showing the remaining strategies in the spatial public goods game with cooperators (C), defectors (D), moralists (M) and immoralists (I), after a sufficiently long transient time.} Initially, each of the four strategies occupies 25\% of the sites of the square lattice, and their distribution is uniform in space. However, due to their evolutionary competition, two or three strategies die out after some time. The finally resulting state depends on the synergy $r$ of cooperation, the punishment cost $\gamma$, and the punishment fine $\beta$.
The displayed phase diagrams are for (a) $r=2.0$, (b) $r=3.5$, and (d) $r=4.4$. (d) Enlargement of the small-cost area for $r=3.5$. Solid separating lines indicate that the resulting fractions of all strategies change continuously with a modification of the model parameters $\beta$ and $\gamma$, while broken lines correspond to discontinuous changes. All diagrams show that cooperators cannot stop the spreading of moralists, if only the fine-to-cost ratio is large enough. Furthermore, there are parameter regions where moralist can crowd out cooperators in the presence of defectors. Note that the spreading of moralists is extremely slow and follows a voter model kind of dynamics \cite{Dornic}, if their competition with cooperators occurs in the absence of defectors. Therefore, computer simulations had to be run over extremely long times (up to $10^7$ iterations for a systems size of $400 \times 400$). For similar reasons, a small level of strategy mutations (which permanently creates a small number of strategies of all kinds, in particular defectors) can largely accelerate the spreading of moralists in the M phase, while it does not significantly change the resulting fractions of the four strategies \cite{accelerate}. The existence of immoralists is usually not relevant for the outcome of the evolutionary dynamics. Apart from a very small parameter area, where immoralists and moralists coexist, immoralists are quickly extinct. Therefore, the 4-strategy model usually behaves like a model with the three strategies C, D, and M only. As a consequence, the phase diagrams for the latter look almost the same like the ones presented here \cite{three}.}
\label{fig1}
\end{figure}

\begin{figure}[!ht]
\begin{center}
\includegraphics[width=5.5in]{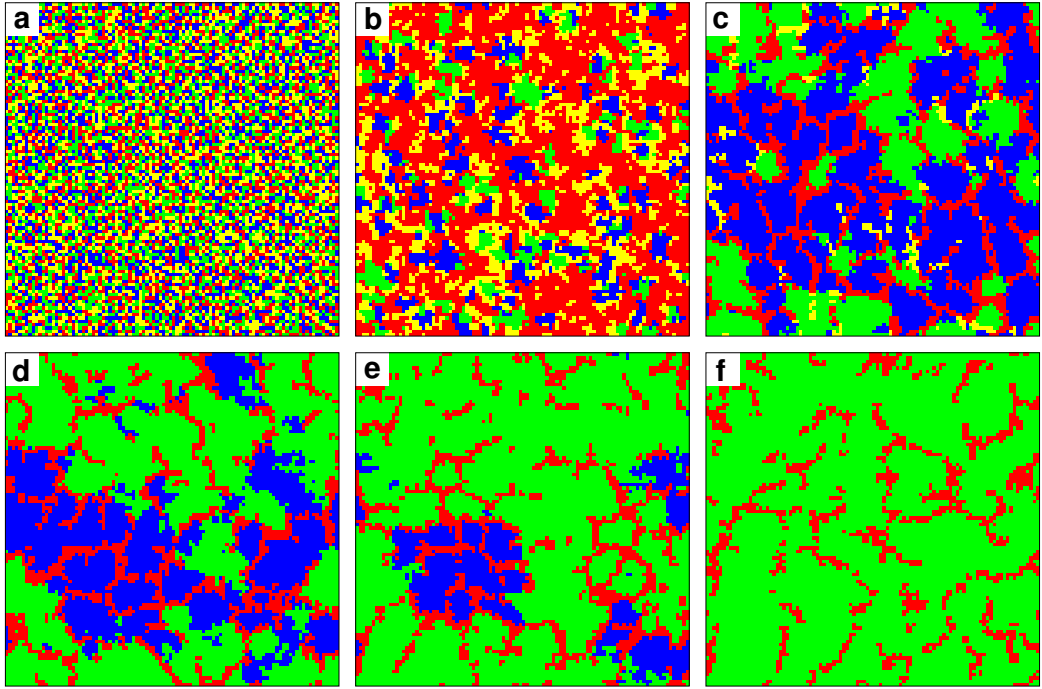}
\end{center}
\caption{{\bf Elimination of second-order free-riders (non-punishing cooperators) in the spatial public goods game with costly punishment for $r=4.4$, $\beta=0.1$, and $\gamma=0.1$.} (a) Initially, at time $t=0$, cooperators (blue), defectors (red), moralists (green) and immoralists (yellow) are uniformly distributed over the spatial lattice. (b) After a short time period (here, at $t=10$), defectors prevail. (c) After 100 iterations, immoralists have almost disappeared, and cooperators prevail, since cooperators earn high payoffs when organized in clusters. (d) At $t=500$, there is a segregation of moralists and cooperators, with defectors in between. (e) The evolutionary battle continues between cooperators and defectors on the one hand, and defectors and moralists on the other hand (here at $t=1000$). (f) At $t=2000$, cooperators have been eliminated by defectors, and a small fraction of defectors survives among a large majority of moralists. Interestingly, each strategy (apart from I) has a time period during which it prevails, but only moralists can maintain their majority. While moralists perform poorly in the beginning, they are doing well in the end. We refer to this as the ``who laughs last laughs best'' effect.}
\label{fig2}
\end{figure}

\begin{figure}[!ht]
\begin{center}
\includegraphics[width=5.5in]{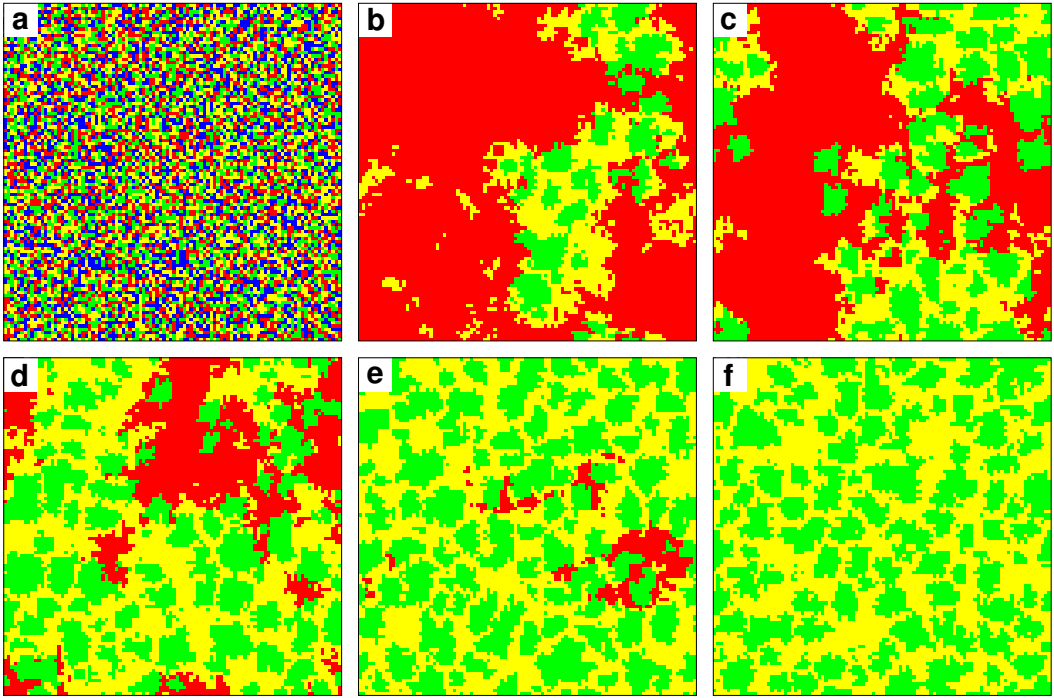}
\end{center}
\caption{{\bf Coexistence of moralists and immoralists for $r=3.5$, $\beta=0.12$, and $\gamma=0.005$, supporting the occurrence of individuals with `double moral standards' (who punish defectors, while defecting themselves).} (a) Initially, at time $t=0$, cooperators (blue), defectors (red), moralists (green) and immoralists (yellow) are uniformly distributed over the spatial lattice. (b) After 250 iterations, cooperators have been eliminated in the competition with defectors (as the synergy effect $r$ of cooperation is not large enough), and defectors are prevailing. (c--e) The snapshots at $t=760$, $t=2250$, and $t=6000$ show the interdependence of moralists and immoralists, which appears like a tacit collaboration. It is visible that the two punishing strategies win the struggle with defectors by staying together. On the one hand, due to the additional punishment cost, immoralists can survive the competition with defectors only by exploiting moralists. On the other hand, immoralists support moralists in fighting defectors. (f) After 12000 iterations, defectors have disappeared completely, leading to a coexistence of clusters of moralists with immoralists.}
\label{fig3}
\end{figure}

\begin{figure}[!ht]
\begin{center}
\includegraphics[width=3.6in]{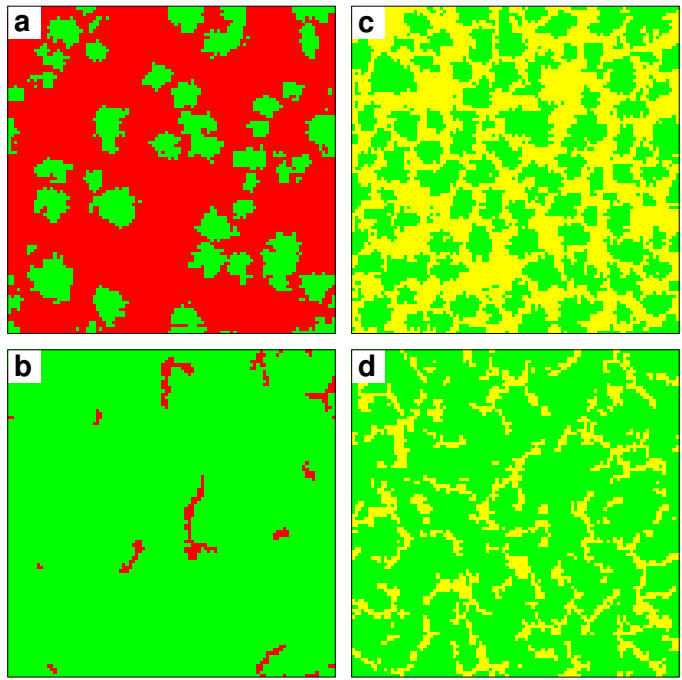}
\end{center}
\caption{{\bf Dependence of cluster shapes on the punishment fine $\beta$ in the stationary state, supporting an adaptive balance between the payoffs of two different strategies at the interface between competing clusters.} Snapshots in the top row were obtained for low punishment fines, while the bottom row depicts results obtained for higher values of $\beta$. (a) Coexistence of moralists and defectors for a synergy factor $r=3.5$, punishment cost $\gamma =0.20$, and punishment fine $\beta =0.25$. (b) Same parameters, apart from $\beta = 0.4$. (c) Coexistence of moralists and immoralists for $r=3.5$, $\gamma=0.05$, and  $\beta=0.12$. (d) Same parameters, apart from $\beta = 0.25$. A similar change in the cluster shapes is found for the coexistence of cooperators and defectors, if the synergy factor $r$ is varied.}
\label{add}
\end{figure}

\begin{figure}[!ht]
\begin{center}
\includegraphics[width=4.5in]{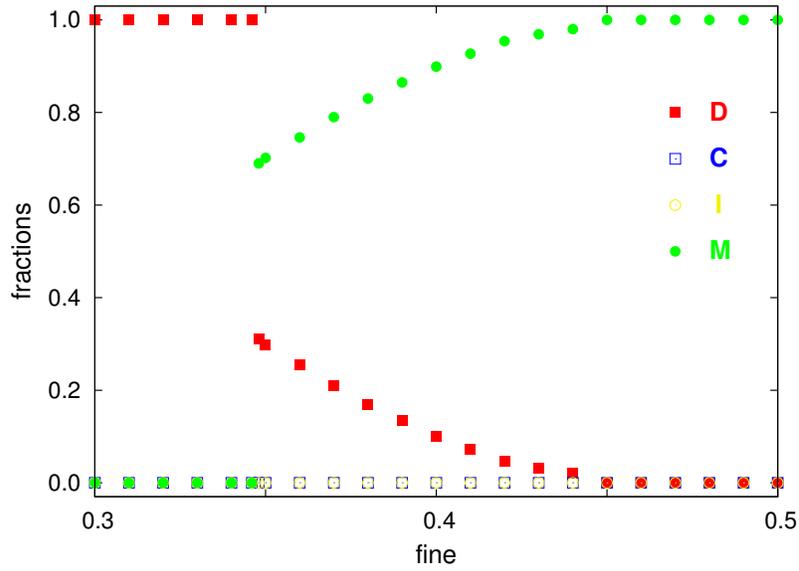}
\end{center}
\caption{{\bf Resulting fractions of the four strategies C, D, I, and M, for random regular graphs as a function of the punishment fine $\beta$.} The graphs were constructed by rewiring links of a square lattice of size $400 \times 400$ with probability $Q$, thereby preserving the degree distribution (i.e. every player has 4 nearest neighbors) \cite{rrg}. For small values of $Q$, small-world properties result, while for $Q\rightarrow 1$, we have a random regular graph. By keeping the degree distribution fixed, we can study the impact of randomness in the network structure independently of other effects. An inhomogeneous degree distribution can further promote cooperation \cite{24}. The results displayed here are averages over 10 simulation runs for the model parameters $r=3.5$, $\gamma=0.05$, and $Q=0.99$. Similar results can be obtained also for other parameter combinations.}
\label{add1}
\end{figure}

\end{document}